\documentclass[aps,prl,twocolumn]{revtex4-1}
\usepackage{amsmath,amssymb,color}
\usepackage{graphicx}
\usepackage[hidelinks]{hyperref}
\begin{document}
\begin{flushright}
    YITP-25-102
    \end{flushright}

\title{Violating Weak Cosmic Censorship in AdS$_3$ via Gedanken Experiment}

\author{Masaya Amo}
\affiliation{Center for Gravitational Physics and Quantum Information,
Yukawa Institute for Theoretical Physics,
Kyoto University, Kyoto 606-8502, Japan}

\begin{abstract}
The weak cosmic censorship conjecture (wCCC) might not be as robust as one would expect. We present a potential counterexample to the wCCC in a rotating three-dimensional anti-de Sitter (AdS$_3$) spacetime.
We extend the parameter range of the quantum corrected BTZ (qBTZ) black hole. 
We examine the infall of test particles with large orbital angular momentum into an extremal black hole. 
Our findings indicate that, under certain conditions---such as neglecting the self-force effects---the event horizon can be destroyed, leading to a violation of the wCCC.
The results suggest that this solution should either be excluded from the applicability of the wCCC or should be examined with self-force effects incorporated.
\end{abstract}

\maketitle
\section{Introduction}

The \textit{weak cosmic censorship conjecture} (wCCC), originally proposed by Penrose~\cite{Penrose:1969pc}, posits that singularities arising from gravitational collapse are always concealed within event horizons, thereby preserving the predictability of spacetime for observers staying outside these horizons. This conjecture is fundamental to general relativity, ensuring that naked singularities do not disrupt the deterministic nature of physical laws. 

Violations of the wCCC also have a profound impact on gravitational physics beyond the classical regime. Naked singularities would reveal regions where classical gravity breaks down, highlighting the necessity of quantum gravity to describe these extreme regions. Consequently, if an observable naked singularity were to exist, it could provide a window into quantum gravitational phenomena.

The wCCC is known to be intimately connected to the Penrose inequality~\cite{Penrose:1973um}, which provides an upper bound on the area in terms of the mass of the black hole. To be more specific, violations of the Penrose inequality imply the existence of naked singularities~\cite{Penrose:1973um}. Therefore, investigating the violation of the Penrose inequality offers a crucial avenue for testing the wCCC. 

Despite the extensive work on the wCCC, no consensus has emerged on its rigorous statement, to the best of our knowledge. Key open questions include whether the statement should be restricted to asymptotically flat spacetimes or extended to asymptotically AdS or dS spacetimes; whether the conjecture is robust across different spacetime dimensions; and whether we need to restrict the equations of motion for the matter and gravity.

Many investigations have been carried out to test the wCCC in different physical systems.
Wald~\cite{Wald:1974hkz} initiated a series of thought experiments, showing that the infall of test particles into an extremal Kerr-Newman black hole would not give rise to a naked singularity. 
This pioneering work spurred extensive research into the possibility of violating the wCCC~\cite{Hubeny:1998ga,Jacobson:2009kt,Gao:2012ca}, including studies in higher-dimensional spacetimes~\cite{Bouhmadi-Lopez:2010yjy,Rocha:2014jma,Revelar:2017sem,An:2017phb},
with the cosmological constant~\cite{Zhang:2013tba,Jiang:2023xca}, including of quantum effects~\cite{Matsas:2007bj,Matsas:2009ww,Richartz:2011vf}, and for multiply charged black holes~\cite{Izumi:2024rge}. Notably, some of these studies obtained results suggesting violations of the wCCC due to the neglect of self-force effects. See also Refs.~\cite{Gregory:1993vy,Horowitz:2016ezu,Crisford:2017zpi,Eperon:2019viw} for other scenarios of counterexamples to the wCCC.

In particular, in the Kerr-Newman spacetime, studies that account for self-force effects appropriately have shown that these effects may prevent overspinning or overcharging, reinforcing the wCCC~\cite{Hod:2002pm,Hod:2008zza,Barausse:2010ka,Isoyama:2011ea,Zimmerman:2012zu,Colleoni:2015afa,Colleoni:2015ena}.
Sorce and Wald~\cite{Sorce:2017dst} successfully showed, using second-order variational/perturbative methods  developed in Refs.~\cite{Iyer:1994ys,Iyer:1995kg,Gao:2001ut,Hollands:2012sf}, that prior test-particle indications of violations do not survive once self-force/second-order effects are included. See also Ref.~\cite{Yoshida:2024txh}, which confirmed that the wCCC is also robust for the Reissner-Nordstr\"{o}m-de Sitter black hole based on Lagrangian methods.

In three-dimensional gravity, the Ba\~nados-Teitelboim-Zanelli (BTZ) black hole~\cite{Banados:1992wn,Banados:1992gq} serves as a fundamental solution for exploring gravitational dynamics and thermodynamics. The BTZ black hole has played a pivotal role in understanding the quantum aspects of black holes, particularly within the framework of the AdS/CFT correspondence~\cite{Maldacena:1997re}.
Rocha and Cardoso~\cite{Rocha:2011wp} showed that the BTZ black hole cannot be spun beyond its extremal limit within the test-particle approximation.

Quantum modifications to the BTZ black hole, constructed via braneworld holography, have led to the construction of the quantum BTZ (qBTZ) black hole~\cite{Emparan:1999fd,Emparan:2002px,Emparan:2020znc}.
This accounts for the exact backreaction of quantum conformal fields, providing a more intricate framework for investigating quantum-related features within gravitational systems. Recent research has delved into various facets of the qBTZ black hole, including its thermodynamic properties~\cite{Frassino:2022zaz,Johnson:2023dtf,Frassino:2023wpc,HosseiniMansoori:2024bfi,Wu:2024txe,Xu:2024iji} and a test against the wCCC~\cite{Frassino:2024fin}, which found results consistent with the wCCC.

The thermodynamic properties of the qBTZ black hole have provided significant insights. 
A key relationship between the thermodynamic volume and entropy
was investigated in Refs.~\cite{Frassino:2022zaz,Frassino:2024bjg}, where the reverse isoperimetric inequality (RII) conjecture~\cite{Cvetic:2010jb} is  $\mathcal{R}\geq 1$ with
\begin{align}
\mathcal{R} &\equiv \left(\frac{V}{V_0}\right)^{1/(D-1)} \left( \frac{A_0}{4 S} \right)^{1/(D-2)}\nonumber\\
&= 1 + \frac{\ell_3 \nu}{r_+ x_1} + \mathcal{O}(\nu^2, \tilde{a}^2) \label{eq:RII}
\end{align}
for the qBTZ black hole. Here, $D$ ($=3$ in this case) is the spacetime dimension, \(V\) denotes the thermodynamic volume, \(S\) is the entropy,
$V_0\;(=\pi)$ and $A_0\;(=2\pi)$ are the volume and area of unit surface of constant $t$ and $r$ in the flat spacetime, \(\ell_3\) is the AdS\(_3\) radius, \(\nu\) is a backreaction parameter, \(r_+\) is the horizon radius, \(\tilde{a}\) is the rotation parameter, and \(x_1\) indicates the location of the rotation axis in the higher-dimensional holographically dual spacetime. 
The RII was found to hold in a variety of spacetimes, such as the Kerr-AdS spacetime~\cite{Cvetic:2010jb}, the spacetime with AdS C-metric~\cite{Gregory:2019dtq}, and higher-dimensional spacetimes with a black ring~\cite{Altamirano:2014tva}.
See also Ref.~\cite{Frassino:2024bjg} for the recently proposed quantum RII.

The RII conjecture imposes an upper bound on the entropy determined by the thermodynamic volume, similar to the Penrose inequality.
In Ref.~\cite{Amo:2023bbo}, the RII conjecture was refined to include the effect of angular momentum, analogously to the refined Penrose inequality, which highlighted the analogy between the RII and the Penrose inequality.
Recalling that the violation of the Penrose inequality implies the violation of the wCCC, it is natural to expect that violating the RII could likewise provide a counterexample to the wCCC.

In this letter, we probe the wCCC in AdS$_3$ spacetime by extending the qBTZ parameter space to \(x_1<0\), which may lead to the violation of the entropy bound in Eq.~\eqref{eq:RII}.
We carry out an analysis of extremal black holes perturbed by test particles of large orbital angular momentum to assess whether the wCCC can be violated.
Our approach follows the method of Wald's gedanken experiments~\cite{Wald:1974hkz} for the Kerr-Newman metric and extends the analysis of Ref.~\cite{Frassino:2024fin} to $x_1<0$.
Note that we do not assume any higher-dimensional holographically dual spacetime in our new extended metric.

\section{Metric Construction for Original Quantum BTZ}

Here, we briefly review the qBTZ solution constructed in Ref.~\cite{Emparan:2020znc}. To explore its properties, we begin with the stationary AdS C-metric~\cite{Plebanski:1976gy}, which serves as the bulk spacetime from which the rotating qBTZ metric is derived.
The AdS C-metric is given by
\begin{equation}
\begin{split}
    ds^2=& \frac{\ell^2}{(\ell + r x)^2} \biggl[  -\frac{H_C(r)}{\Sigma(r,x)}\left( dt + a x^2 d\phi \right)^2 + \frac{\Sigma(r,x)}{H_C(r)} dr^2 \\
    & + r^2 \left( \frac{\Sigma(r,x)}{G(x)} dx^2 + \frac{G(x)}{\Sigma(r,x)}\left( d\phi - \frac{a}{r^2} dt \right)^2 \right) \biggr],
\end{split}
\label{eq:AdS4Cmetric}
\end{equation}
where
\begin{subequations}
\begin{align}
    H_C(r) &= r^2 \left( \frac{1}{\ell_4^2} - \frac{1}{\ell^2} \right) + \kappa - \frac{\mu \ell}{r} + \frac{a^2}{r^2}, \label{eq:H_C} \\
    \Sigma(r,x) &= 1 + \frac{a^2 x^2}{r^2},\\
    G(x) &= 1 - \kappa x^2 - \mu x^3 + a^2 \left( \frac{1}{\ell_4^2} - \frac{1}{\ell^2} \right) x^4. \label{eq:G_x}
\end{align}
\end{subequations}
Here, $\kappa$ is $\pm1$ or 0~\footnote{We do not discuss the case with $\kappa=0$ separately since it is obtained as a limit $x_1\to0$ of the case with either $\kappa=\pm1$ (see Ref.~\cite{Emparan:2020znc}).}, $\ell_4$ is the AdS radius of the bulk, \(a\) is the rotation parameter, and the parameter \(\mu\) is associated with the four-dimensional black hole mass \(M_4\) through the relation $M_4=\ell\mu/2$.
Here, we take $\mu$ to be positive.
Let \(x_1>0\) denote the smallest positive root of \(G(x)=0\).
Then, the parameter $\mu$ is expressed in terms of $x_1$ and other parameters through Eq.~\eqref{eq:G_x}.

\begin{figure}[htbp]
    \centering
    \includegraphics[width=0.8\linewidth]{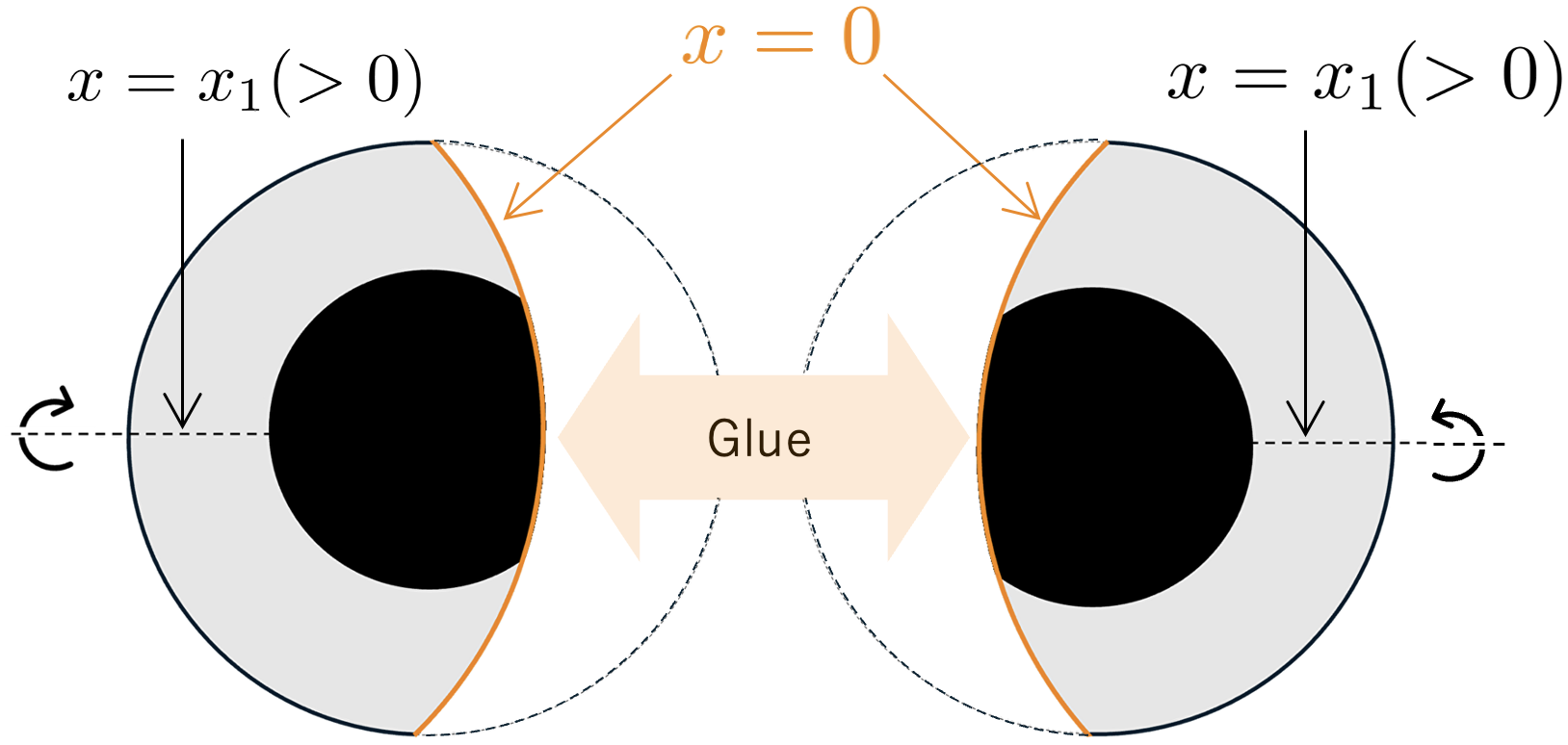}
   \caption{Schematic illustration of the qBTZ configuration. Two copies of the bulk region $0\leq x\leq x_1$ are joined along the brane at $x=0$, resulting in a configuration with $\mathbb{Z}_2$ symmetry.
   }
    \label{fig:patch1}
\end{figure}

The brane is located at \(x=0\), and the coordinate \(x\) is restricted to \(0\le x\le x_1\), ensuring
\(G(x)\ge0\) and preserving the Lorentzian signature.
Two identical copies of this spacetime are joined along the brane at \(x=0\). 
The brane tension is $\tau=1/(2\pi G_4\ell)\,.$
Figure~\ref{fig:patch1} illustrates this setup.
Projecting the bulk metric onto the brane, we obtain the metric of the rotating qBTZ black hole~\cite{Emparan:2020znc}, which is a solution of an effective three-dimensional gravitational field equation with higher curvature corrections
\begin{equation}\label{eq:EOM}
G_{\mu\nu}+\Lambda_3 g_{\mu\nu}+({\rm corrections})=8\pi G_3\langle T_{\mu\nu}\rangle,
\end{equation}
where $\Lambda_3$ is the cosmological constant in three dimensions, and the higher curvature corrections are of order $\mathcal{O}(\nu^{2})$.
See Ref.~\cite{Bueno:2022log} for the detailed expression of higher curvature terms.
The nonzero components of the rotating qBTZ metric are given by~\footnote{We adopt the same notation for the three-dimensional coordinates as in Ref.~\cite{Frassino:2024fin} to facilitate comparison, which differs from that in Ref.~\cite{Emparan:2020znc}.}
\begin{subequations}
\begin{align}
    g_{tt} &= -\left( \frac{r^2}{\ell_3^2} - 8 \mathcal{G}_3 M_3 - \frac{\mu \ell \Delta^2}{\hat{r}} \right), \label{eq:g_tt}\\
    g_{\phi\phi} &= r^2 + \ell_3^2 \, \frac{\mu \ell \tilde{a}^2 \Delta^2}{\hat{r}}, \\
    g_{t\phi} &= -4 \mathcal{G}_3 J_3 \left( 1 + \frac{\ell}{\hat{r} x_1} \right), \\
    g_{rr} &= \left[ \frac{r^2}{\ell_3^2} - 8 \mathcal{G}_3 M_3 + \frac{(4 \mathcal{G}_3 J_3)^2}{r^2} - \mu \ell (1 - \tilde{a}^2)^2 \Delta^4 \, \frac{\hat{r}}{r^2} \right]^{-1}, \label{eq:g_rr}
\end{align}
\end{subequations}
where $\ell_3$ is the AdS$_3$ radius and is related to $\ell_4$ and $\ell$ via $\ell_3^{-2}=\ell_4^{-2}-\ell^{-2}$.
The parameter $\tilde{a}$ is defined as
$\tilde{a}^2\equiv a^2 x_1^4/\ell_3^2$.
In the three-dimensional metric, the parameter
\(\nu\equiv\ell/\ell_3\) characterizes the strength of quantum backreaction effects; smaller values of $\nu$ correspond to smaller backreaction.
The (classical) BTZ metric is obtained by taking $\ell\to0$.
The constant \(\mathcal{G}_3=(\ell_4/\ell)\,G_3\) denotes the renormalized Newton's constant incorporating the higher curvature effects. Parameters \(M_3\) and \(J_3\) represent the mass and angular momentum on the brane and can be expressed as
\begin{subequations}
    \begin{align}
        M_3 &= \frac{1}{2 \mathcal{G}_3} \frac{-\kappa x_1^2 + \tilde{a}^2 (4 - \kappa x_1^2)}{(3 - \kappa x_1^2 - \tilde{a}^2)^2}, \label{eq:M} \\
        J_3 &= \frac{\ell_3}{\mathcal{G}_3} \frac{\tilde{a} (1 - \kappa x_1^2 + \tilde{a}^2)}{(3 - \kappa x_1^2 - \tilde{a}^2)^2}. \label{eq:J}
    \end{align}
    \label{eq:parameters}
    \end{subequations}
The parameter \(\Delta\) is determined to eliminate a conical singularity at \(x=x_1\) in the bulk and is given by
\begin{equation}
\Delta=\frac{2}{|G'(x_1)|}.\label{eq:conical}
\end{equation}
We restrict \(\tilde{a}\) to satisfy \(\tilde{a}^2<1\) in order to avoid closed timelike curves. Below, we concentrate on the case \(0<\tilde{a}<1\).
The unphysical radial coordinate $\hat{r}$ is given by
\begin{equation}
r^2=(1 - \tilde{a}^2) \Delta^2 \hat{r}^2 + r_S^2,
\label{eq:rtor}
\end{equation}
where $r_S$ is 
\begin{equation}
r_S=\ell_3 \frac{2 \tilde{a} \sqrt{2 - \kappa x_1^2}}{3 - \kappa x_1^2 - \tilde{a}^2}.
\label{eq:rS} 
\end{equation}
Note that $r$ is restricted to $r\ge r_S$, ensuring that \(\hat{r}\) remains real.
The renormalized holographic CFT stress tensor in Eq.~\eqref{eq:EOM} is calculated in Ref.~\cite{Emparan:2020znc}, yielding
\begin{subequations}
        \begin{align}
        8\pi G_3\langle T_{tt} \rangle &= -\frac{\Delta^{3}\sqrt{1-\tilde{a}^{2}}\,(1+2\tilde{a}^{2})\,\mu\,\nu}{2\,r\,\ell_{3}}+ \mathcal{O}(r^{-3}, \nu^2),\label{eq:EMTtt}\\
        8\pi G_3\langle T_{\phi t} \rangle &=\frac{3\Delta^{3}\tilde{a}\sqrt{1-\tilde{a}^{2}}\,\mu\,\nu}{2\,r}+ \mathcal{O}(r^{-3}, \nu^2),\\
        8\pi G_3\langle T_{\phi \phi} \rangle &=-\frac{\Delta^{3}\sqrt{1-\tilde{a}^{2}}\,(2+\tilde{a}^{2})\,\ell_3 \,\mu\,\nu}{2\,r}+ \mathcal{O}(r^{-3}, \nu^2),\\
        8\pi G_3\langle T_{rr} \rangle &=
        \frac{\Delta^{3}(1-\tilde{a}^{2})^{3/2}\,\ell_{3}^{\,3}\,\mu\,\nu}{2\,r^{5}}+ \mathcal{O}(r^{-7}, \nu^2).\label{eq:EMTrr}
        \end{align}
    \end{subequations}

We note that the three-dimensional coordinates in Eqs.~\eqref{eq:g_tt}--\eqref{eq:g_rr} are not simply induced from the four-dimensional coordinates in Eq.~\eqref{eq:AdS4Cmetric}.
To ensure that the three-dimensional angular coordinate $\phi$ is properly identified such that $\phi\sim\phi+2\pi$, and to eliminate any residual frame dragging at infinity, a coordinate transformation is necessary~\cite{Emparan:2020znc}. 
However, for the purposes of this letter, we present only the final form of the metric without delving into the details of the coordinate transformation.

\section{Our Metric Construction}

In this work, we consider the metric that has the same expressions as Eqs.~\eqref{eq:g_tt}--\eqref{eq:rS} but with $\mu<0$ and $x_1<0$, where $x_1$ is defined as the largest negative root of \(G(x)=0\), rather than the smallest positive root. 
The coordinate domain is $x_1\leq x\leq 0$.
We may regard $x_1$ as a parameter of the solution instead of $\mu$. 
The stress tensor has the same form as Eq.~\eqref{eq:EMTtt}--\eqref{eq:EMTrr}, which suggests that energy conditions, including averaged ones, are more favorable in \(x_1<0\) (\(\mu<0\)) than \(x_1>0\) (\(\mu>0\)) at least in the far zone, at the leading order in \(\nu\).

Note that we assume the metric but do not assume a specific gravitational equation of motion. For example, one can assume Eq.~\eqref{eq:EOM} as the equations of motion with stress tensor Eq.~\eqref{eq:EMTtt}--\eqref{eq:EMTrr}.
Although the relation between $\mathcal{G}_3$ and $G_3$ depends on higher-curvature terms (with $\mathcal{G}_3=G_3$ if no such terms are present), our wCCC criterion---based on the sign of $\min g^{rr}$, as we will see later---is insensitive to this redefinition. Hence, our conclusions do not hinge on the precise $\mathcal{G}_3–G_3$ relation.

\section{Gedanken Experiment Setup and Analysis}

In this section, we investigate the possibility of violating the wCCC by attempting to overspin an extremal rotating black hole using a test particle with large orbital angular momentum. We follow the method introduced by Wald~\cite{Wald:1974hkz} and extend the analysis in Ref.~\cite{Frassino:2024fin} for $x_1>0$ to the case where $x_1<0$. In Ref.~\cite{Frassino:2024fin}, it was shown that the first-order contribution in infinitesimal $\delta M_3$ does not affect the final outcome, and they proceeded to solve equations for a finite perturbation $\delta M_3=0.01M_3$. 
Here, we extend the analysis to the $x_1<0$ case, focusing on the second-order (leading non-vanishing) contribution in an infinitesimal \(\delta M_3\), thereby ensuring the validity of our discussion within the perturbative regime and minimizing the backreaction effects.

Our physical procedure is outlined as follows.

\smallskip

1. \textbf{Initial Extremal Black Hole:} We consider an extremal rotating black hole with $(M_3,J_3)$. 

2. \textbf{Infall of Test Particle:} We introduce a test particle with mass $\delta M_3 (>0)$ and large orbital angular momentum $\delta J_3 (>0)$, which is absorbed by the black hole.

3. \textbf{Merger Process:} The black hole and the test particle undergo a merging process, while conserving the total energy and angular momentum.

4. \textbf{Final State:} After the merger, the black hole's parameters become $(M_3+\delta M_3,J_3+\delta J_3)$. 

\smallskip

We make an assumption that the final state is described by the metric in Eqs.~\eqref{eq:g_tt}--\eqref{eq:g_rr}. 
The parameters are $x_1$, $\nu$, $\ell_3$, $\tilde{a}$, and $\kappa$. 
For the initial state, $x_1$ is determined by the other parameters due to the extremality.

We analyze whether the final state corresponds to a black hole or a naked singularity in the following steps:

\smallskip

1. \textbf{Parameter Specification:} We fix parameters for the initial black hole.
Then, we compute the initial mass $M_3$ and angular momentum $J_3$ by Eqs.~\eqref{eq:M} and \eqref{eq:J}.

2. \textbf{Determination of $\delta J_3$:} We consider a test particle that barely falls into the black hole. 
Requiring its geodesic to remain future-directed even on the horizon, $\delta J_3$ can be taken as large as
\begin{equation}
    \delta J_3=-\left.\frac{g_{\phi\phi}}{g_{t\phi}}\right|_\mathcal{H} \delta M_3 - \varepsilon,
\end{equation}
where $\mathcal{H}$ is the horizon. $\varepsilon>0$ is of sufficiently higher order in $\delta M_3$, which accounts for the fact that without this correction, the particle would spiral asymptotically toward the horizon rather
than cross it. 
However, details of this term do not contribute to our final result at the leading order. Thus, we set $\varepsilon=0$ to obtain the leading-order result.

3. \textbf{Final Parameters:} Using the final mass \(M_3+\delta M_3\) and angular momentum \(J_3+\delta J_3\), 
the final values of \(\tilde{a}\) and \(x_1\) are obtained by solving Eqs.~\eqref{eq:M} and \eqref{eq:J}.

4. \textbf{Existence of Horizon:} We compute \(g^{rr}(r)\) for the final state and determine its minimum value for \(r>0\), written as \(\min\,g^{rr}(r)\). Since \(\lim_{r\to\infty}g^{rr}(r)=\infty\), a positive minimum value implies that \(g^{rr}(r)>0\) for all \(r>0\). Then, \(\min\,g^{rr}(r)>0\) indicates the absence of a horizon and the formation of a naked singularity, suggesting a violation of the wCCC under the assumptions made.

\smallskip

We note that, once $\nu$, $\tilde{a}$, and $\kappa$ are fixed, \(\ell_3\) enters the metric solely through \(\ell_3/r\). Consequently, the specific value of \(\ell_3\) does not affect the final result of the wCCC.

\section{Results}

\begin{figure}[htbp]
    \centering
    \includegraphics[width=0.91\linewidth]{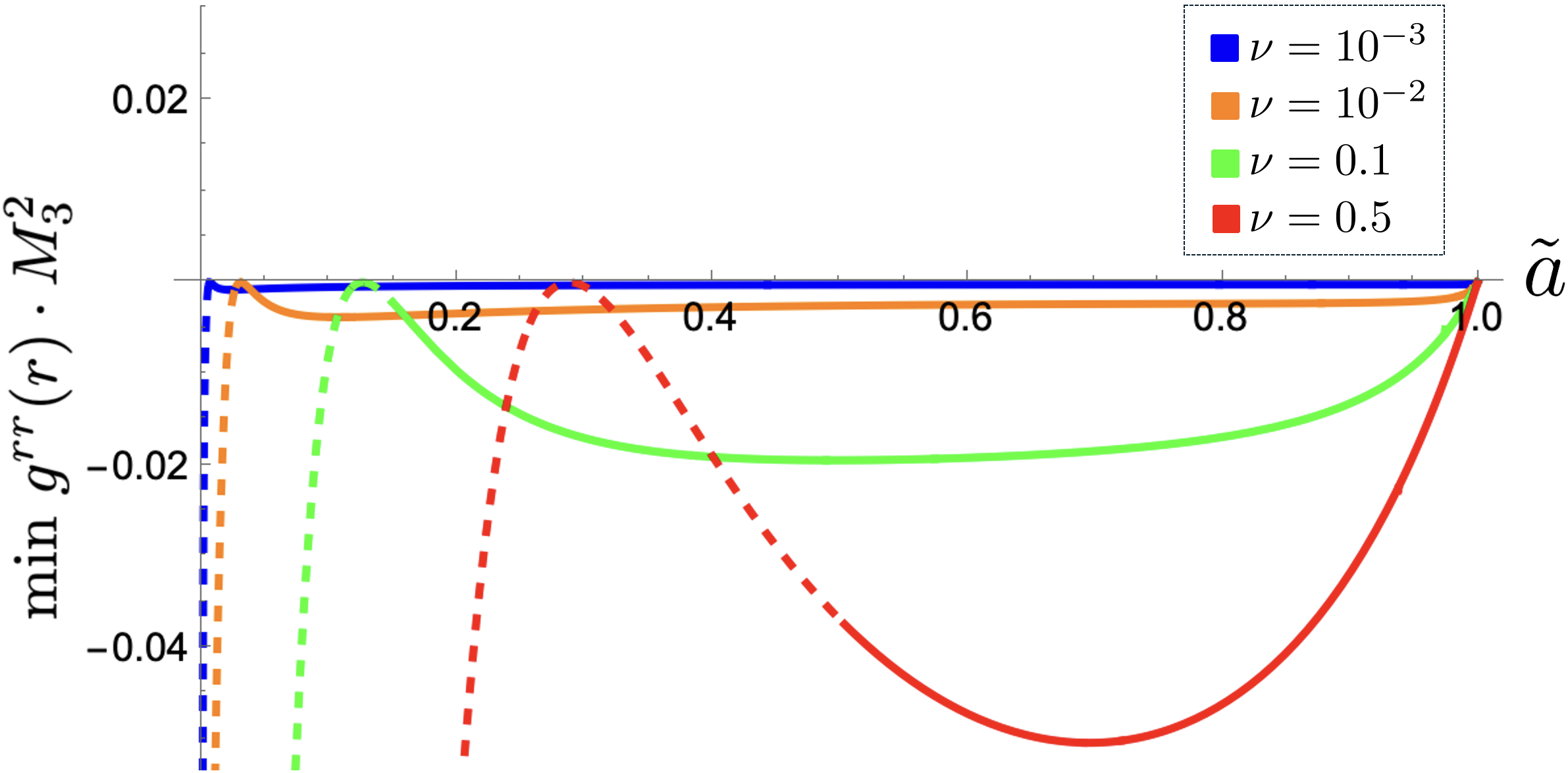}
    \caption{The minimum of $g^{rr}(r)$ times $M_3^{\,2}$ as a function of \(\tilde{a}\) for \(x_1>0\) at the second order in \(\delta M_3\). 
    Dashed lines indicate $\kappa=1$, while solid lines indicate $\kappa=-1$.
    The resulting non-positive values indicate that the event horizon is not destroyed, thereby upholding the wCCC. These findings are consistent with the results of Ref.~\cite{Frassino:2024fin}, in which the minimum of \(g^{rr}(r)\) is calculated for $\delta M_3=0.01M_3$.
    }
    \label{fig:positive_x1}
\end{figure}

\begin{figure}[htbp]
    \centering
    \includegraphics[width=0.9\linewidth]{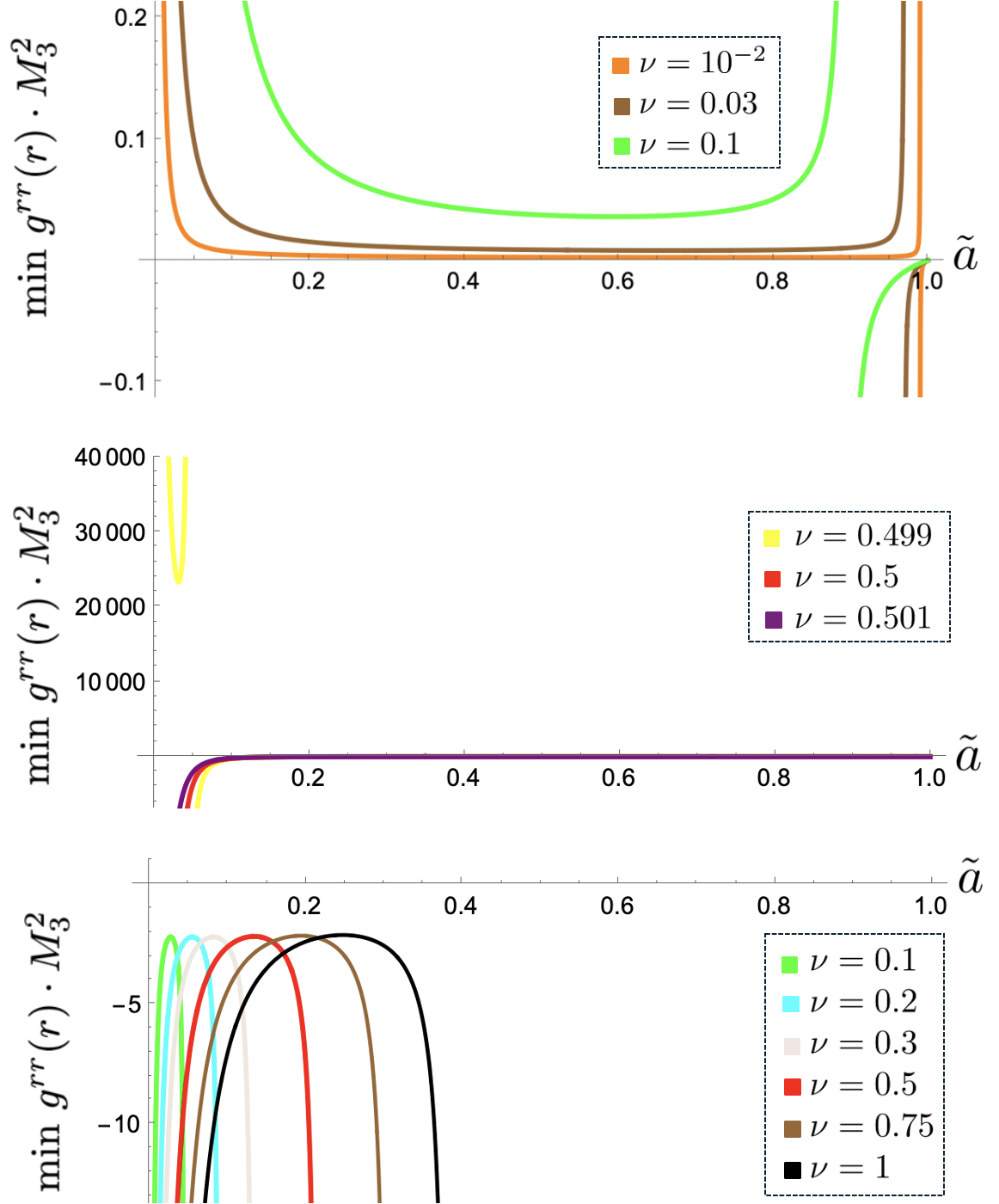}
    
    \caption{Plots of $\min\,g^{rr}(r)\cdot M_3^2$ as a function of \(\tilde{a}\) for \(x_1<0\) at the second order in \(\delta M_3\). (Top) Curves for \(\nu=10^{-2},0.03\), and \(0.1\) with $\kappa=-1$. Positive values indicate the disappearance of the horizon, suggesting a violation of the wCCC. 
    (Middle) Curves for \(\nu=0.499,0.5\), and \(0.501\) with $\kappa=-1$. These plots show that for \(\nu \geq 0.5\), \(\min\,g^{rr}(r)\) remains non-positive for any value of \(\tilde{a}\), indicating no violation of the wCCC. (Bottom) Curves for \(\nu=0.1,0.2,0.3,0.5,0.75\), and \(1\) with $\kappa=+1$. For these parameters, no violation was observed.}
    \label{fig:negative_x1}
\end{figure}

Here, we present the results of our analysis.
For comparison, we first calculate the leading (second-order) contributions in \(\delta M_3\) to the \(\min\,g^{rr}(r)\cdot M_3^2 \) for \(x_1>0\) (Fig.~\ref{fig:positive_x1}), which closely match the findings of Ref.~\cite{Frassino:2024fin}.
The minimum remains negative after the perturbation, indicating that the horizon does not disappear.

\begin{figure}[htbp]
    \centering
    \includegraphics[width=0.83\linewidth]{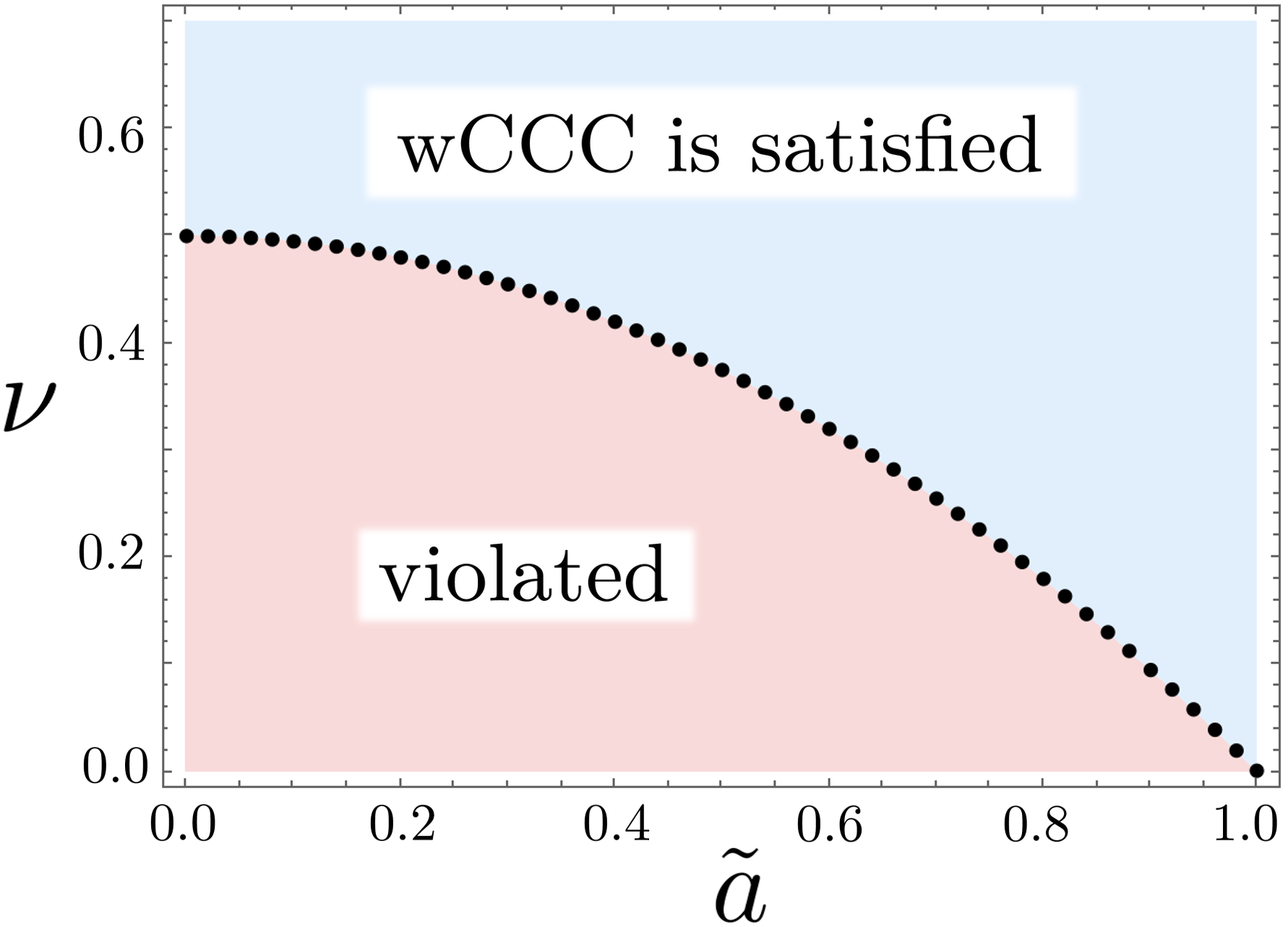}
    \caption{Parameter regions illustrating the (in)validity of the wCCC for \(\kappa=-1\) and  \(x_1<0\). Black dots denote boundary points for values of \(\nu\). No violation was found for $\nu\ge0.5$. 
    Intriguingly, all of the boundary points for $\nu<0.5$ are consistent with the curve \(0.5-\tilde{a}^2/2=\nu\) within differences of order $10^{-16}$ in $\tilde{a}$, which are numerical errors.
    }
    \label{fig:violate_region}
\end{figure}

Let us move on to the case with $x_1<0$. Figure~\ref{fig:negative_x1} illustrates the leading order in $\delta M_3$ of the minimum value of $g^{rr}(r)\cdot M_3^{\,2}$ as a function of $\tilde{a}$. Notably, the minimum of \(g^{rr}(r)\) becomes positive for some parameter sets in the case with $\kappa=-1$, suggesting the disappearance of the horizon. 
This implies a violation of the wCCC under our assumptions. 
It is intriguing that the wCCC-violating region of parameter space is well described by $0.5-\tilde{a}^2/2\ge\nu$ as shown in Fig.~\ref{fig:violate_region}. In particular, we found no violation for $\nu\ge0.5$.
The results for $\kappa=+1$ are consistent with the wCCC as shown in Fig.~\ref{fig:negative_x1}.

\section{Conclusion \& Discussion}

In this letter, we investigated the wCCC in the extended metric of the rotating qBTZ black hole to $x_1<0$. 
Extending the analysis of Ref.~\cite{Frassino:2024fin} for the original qBTZ black hole with finite $\delta M_3$, we considered infinitesimal perturbations up to second order in $\delta M_3$ with a test-particle approximation for the AdS$_3$ black hole with metric given by Eqs.~\eqref{eq:g_tt}--\eqref{eq:g_rr} with $x_1<0$.  Our findings indicate that it may be possible to overspin the black hole to form a naked singularity.

Our results suggest that the violation of the wCCC in this scenario implies one of two possibilities: either (i) the wCCC still holds if we resolve some of the assumptions in the analysis or (ii) the wCCC does not hold for the metric given by Eqs.~\eqref{eq:g_tt}--\eqref{eq:g_rr} with $x_1<0$. As for (i), one of our assumptions is to neglect backreaction effects.
A detailed investigation incorporating self-force effects along the lines of Ref.~\cite{Sorce:2017dst} would be meaningful to determine whether the wCCC remains violated when these effects are properly accounted for.
As for (ii), it may be the case that, if the wCCC becomes a {\it weak cosmic censorship theorem} in the future, the statement of the theorem requires  restrictions to exclude the metric we  considered here. 

One may also ask whether the wCCC holds for \(\nu=\ell/\ell_3<0\) (with \(\ell<0\) and \(\ell_3>0\)). However, simultaneously applying the transformations \(\nu\mapsto-\nu\), \(\ell\mapsto-\ell\), \(x_1\mapsto-x_1\), and \(\mu\mapsto-\mu\) simply reproduces the original metric in three dimensions. In other words, by considering both \(x_1>0\) and \(x_1<0\), all cases are effectively covered by solely analyzing scenarios with \(\nu\ge0\).

Several questions remain open. 
One direction is to generalize the framework to charged qBTZ geometries~\cite{Feng:2024uia,Climent:2024nuj,Bhattacharya:2025tdn}.
Additionally, repeating the gedanken experiment from a near-extremal, rather than exactly extremal black hole, would be important as well. 
In addition, generalizing our result to relax Eq.~\eqref{eq:conical} is left for future work.
We did not delve into the detailed relationship between the RII and the wCCC in this letter because we found no clear relationship as strong as initially expected between them. Their precise relationship is unresolved. 
Finally, clarifying why the parameter region that violates the wCCC is so simply characterized by $0.5-\tilde{a}^2/2\ge\nu$ (Fig.~\ref{fig:violate_region}) remains an important task for future work.

In conclusion, our analysis raises further questions about the validity of the wCCC in AdS$_3$ and highlights the importance of thoroughly examining the assumptions underlying such investigations. 

\section*{Acknowledgements}
The author is grateful to Roberto Emparan, Antonia M.~Frassino, Keisuke Izumi, Shinji Mukohyama, Jorge V. Rocha, Tetsuya Shiromizu, Takahiro Tanaka, and Daisuke Yoshida for valuable discussions, suggestions, and comments.
This work was supported by Grant-in-Aid for JSPS Fellows (No.22J20147 and 22KJ1933).

\bibliography{thebib_CCC}

\end{document}